# Mesoscopic modeling for nucleic acid chain dynamics

M. Sales-Pardo*, R. Guimerà*, A.A. Moreira*, J. Widom$^\sharp$, and L.A.N. Amaral*

*Department Chemical and Biological Engineering, Northwestern University

2145 Sheridan Road, Evanston, IL 60208, USA

$^\sharp$Department of Biochemistry, Molecular Biology, and Cell Biology,

and Department of Chemistry, Northwestern University

2153 Sheridan Road, Evanston, IL 60208, USA.

## Abstract

To gain a deeper insight into cellular processes such as transcription and translation, one needs to uncover the mechanisms controlling the configurational changes of nucleic acids. As a step toward this aim, we present here a novel mesoscopic-level computational model that provides a *new window* into nucleic acid dynamics. We model a single-stranded nucleic as a polymer chain whose monomers are the nucleosides. Each monomer comprises a bead representing the sugar molecule and a pin representing the base. The bead-pin complex can rotate about the backbone of the chain. We consider pairwise stacking and hydrogen-bonding interactions. We use a modified Monte Carlo dynamics that splits the dynamics into translational bead motion and rotational pin motion. By performing a number of tests we first show that our model is physically sound. We then focus on the study of a the kinetics of a DNA hairpin—a single-stranded molecule comprising two complementary segments joined by a non-complementary loop—studied experimentally. We find that results from our simulations agree with experimental observations, demonstrating that our model is a suitable tool for the investigation of the hybridization of single strands.

## I. INTRODUCTION

Some of the most challenging questions in biochemistry—such as determining RNA secondary structure starting from sequence alone [1, 2] or identifying the dynamic mechanism responsible for the slow folding of the molecule into its catalytic structure [3, 4]—concern the mesoscopic behavior of nucleic-acid chains. The understanding of the configurational changes of nucleic acids is a key step if one wishes to control cellular processes such as transcription or translation. In addition, the configurational dynamics of single-stranded nucleic acids is also relevant to microarray experiments: The expression level assigned to a given gene is related to the hybridization of a labeled nucleic-acid chain (the probe) to another nucleic-acid chain tethered to a glass slide (the target) [5–7]. In microarrays, each gene is represented in 10 to 20 spots. Significantly, the hybridization yields for spots representing the same gene exhibit large fluctuations, posing serious problems for the interpretation of microarray results [8–10]. Understanding the hybridization of target and probe will thus help us in designing more reliable microarrays and in interpreting microarray data.

Nucleic-acid hairpins are likely the least complex system from which to assess mesoscopic properties of single strands. They are also relevant to a number of biologically important phenomena. For example, in RNA, the formation of hairpin structures is believed to be the critical step before the fast folding into the native configuration [11], while, in DNA, hairpin formation is relevant to a number of significant processes such as recombination, transposition, and gene expression [12–14]. For these reasons, hairpins are systems to which experimentalists have devoted much attention [15–20]. Importantly, experimental observations report that, even for short hairpins, the configurational dynamics is complex and strongly affected by sequence.

Here, we develop a mesoscopic-level model which we show can describe the dynamics of single-stranded nucleic acids. In order to validate our model, we study short DNA hairpins— single-stranded nucleic acid chains comprising two complementary "stems" joined by a non-complementary "loop." We show that simulations of the model consistently reproduce predicted melting temperatures. To validate the dynamics, we focus our attention on a DNA hairpin which was extensively studied experimentally by Ansari and co-workers [16, 18] and show that the relaxation rates measured with our model agree with the relaxation rates measured experimentally.

This paper is organized as follows. In Sec. II, we review the existing modeling approaches for DNA. In Sec. III, we describe our model including the basic units, the types of interactions and the implementation of the dynamics. In Sec. IV, we present the results of a number of tests used to validate the model, including the comparison with experimental observations for an extensively studied hairpin. Finally, in Sec.V we present our conclusions.

## II. PRIOR NUCLEIC ACID MODELING

Nucleic acids are linear polynucleotide chains. Each nucleotide comprises a nitrogenous organic base attached to a pentose—a five-carbon sugar—which is also attached to a phosphoric acid. The pentose in DNA is a deoxyribose, while in RNA the pentose is a ribose. The carbon atoms in the pentose are labeled from $1'$, the carbon to which the base is attached, to $5'$, to which the phosphate group is attached. The bases fall onto two groups: The *purines*—adenine (A) and guanine (G)—and the *pyrimidines*—thymine (T), cytosine (C), and uracil (U). The combination of a nucleic base and a pentose is called a nucleoside. A nucleotide is formed by attaching one, two, or tree phosphate groups to a nucleoside.

*Ab initio modeling*—For short time and length scales, researchers typically use *ab initio* models, in which interactions between atoms are calculated by integration of the Schroedinger equation [21–23]. Since the electron orbitals are explicitly considered, this approach is adequate to investigate phenomena involving changes in electronic states such as chemical reactivity and absorption of light.

A weakness of *ab initio* modeling is that it takes into account neither the molecular structure nor solvent or temperature effects. Thus, these methods only describe the zero-temperature gas phase of nucleic acids. Nonetheless, the information obtained from *ab initio* calculations provides the theoretical grounds for the parametrization of more coarse-grained models [24].

*Force-field models*— Due to the complexity of *ab initio* calculations, the use of these models is restricted to single nucleotides or oligonucleotide dimers [25]. To model nucleic acids at larger scales, one can use force-field models [26, 27] in which the DNA molecule is treated as a *classical system* composed of atoms held together by bonds. In these models, the energy of the system is a function only of the position of the atoms.

Force-field models have successfully predicted both static [28, 29] and dynamic [30, 31] structural properties of DNA. However, a serious handicap of this treatment is that the existence of a large number of long-range electrostatic pairwise interactions dramatically increases the duration of the simulations. To overcome this problem, one can truncate the potential, but this leads to the construction of an effective potential that is not necessarily accurate. Nevertheless, a chain with 12 base pairs can be simulated for typically 20 nanoseconds, which is the timescale associated with the rotation of a nucleotide [32].

*Zipper models*—The computational cost of force-field models imposes the need to develop even coarser descriptions in order to model longer timescales or longer chains. To characterize DNA denaturation, a successful approach is to consider a two-dimensional lattice model in which the two strands are bonded by springs and bases oscillate about their equilibrium position [33]. An alternative approach are Ising-like models—which describe double-stranded (ds) DNA as an ensemble of molecule configurations in which bases are either open or closed [34–36]—are quite accurate in predicting equilibrium properties such as the melting temperatures of large chains. Recently these models have been extended by including elasticity terms in order to describe different dynamic aspects observed in the so-called pulling and unzipping experiments of single molecules [19, 37, 38]. However, most of these models do not consider sequence heterogeneity, and even when they do, they do not take into account the sequence dependence of the single-strand contribution.

*Bead models*— A second class of mesoscopic models are the so-called bead models, which are used to study the long-time dynamics of DNA molecules [39–42]. In these models, each DNA single strand is a chain of beads. Each bead represents a rigid part of the nucleotide [43] or the center of mass of bases and backbone groups [40]. Bead models—which successfully reproduce the melting dynamics observed in experiments [40]—typically consider only interactions that affect double-helix stability, neglecting single-strand properties.

*Elastic chains*—To investigate even larger molecules, one has to introduce further simplifications. For instance, to investigate the supercoiling structure of dsDNA in chromosomes, researchers model dsDNA as an elastic chain whose units interact electrostatically [44]. With these models, it is possible to investigate the dynamics of very long chains containing thousands of base pairs for timescales on the order of milliseconds [45]—the timescale associated with site specific recombination processes [46]—as well as the temperature- or torque-induced denaturation of long molecules [47].

## A. Mesoscopic models for single-stranded nucleic acids

Recently, several groups have developed models to investigate the statics and dynamics of single-stranded nucleic acids at mesoscales [10, 48–52]. Most of these models focus on the investigation of a system of great current interest: hairpins, which are single-stranded nucleic acids with two complementary sections linked by a non-complementary loop [15–18]. Hairpins appear in both DNA and RNA and participate in a number of biological processes such as recombination and gene expression mechanisms [46, 53]. All-atom models have also been used to study mesoscopic objects like hairpins [54, 55]. Specifically, Sorin et al. have investigated the configurational dynamics of an RNA hairpin fourteen bases long. Using 40,000 processors, they could simulate the molecule for 500 $\mu s$. This is clearly the largest scales that one can pursue with such models, but, unfortunately, it still falls short for the time scales involved in microarray experiments, which are of the order of seconds or more.

The models proposed for the study of hairpins fall roughly into two categories. In the first category, one finds models whose purpose is to investigate the elastic properties of the hairpin loops [49, 52]. In these models the hairpin has no stem. It is reduced to an homopolymers which represents the loop and whose monomers in the loop can be either stacked or unstacked. For example, in the model of Aalberts et al. [49], the polymer is divided into rigid segments comprised by an equal number of monomers to mimic different stacking strengths, whereas, in the model of Sain et al. [52], the stacking interaction between neighboring monomers is specifically taken into account. While these models are a practical first approach to investigate ring formation of single-stranded nucleic acids, their use is very limited, since the dynamics of hairpins with long stems cannot be investigated.

In the second category, one finds "configurational models." These models are defined on a plane, thus they only consider the secondary structure of the hairpin [10, 48, 50, 51]. Different configurations in these models differ in the sequence of base pairs bonded.

The model proposed by Chen and Dill [48] uses polymer graph theory to compute the entropy associated to the different configurations and uses a multiplicative factor to account for the loss in entropy due to the missing third dimension. The stacking free-energies for the different configurations are computed for each particular sequence using the Turner rules [56]. However, single-stranded regions have no energy contribution. Recently, Zhang and Chen [10] studied the "configurational" dynamics with these types of models by introducing transition rates. The only allowed transitions are those that break or add a base pair to configurations comprising at least two stacked base pairs.

Cocco et al. proposed a similar model to study the unzipping dynamics of the pulling experiments on RNA hairpins. In their model, free-energies are also computed using the Turner Rules and an extra entropic term is assigned to the single-stranded ends of the molecule. The dynamics is implemented by assigning transition rates to the process of braking or adding a single base pair at each time step.

All these models have a common feature: They rely on the zipping/unzipping mechanisms to describe the folding and unfolding of hairpins. This approach has been proved useful to study some aspects of how the dynamics relate to the free-energy landscape. However, since there is no sequence-specific treatment of a single strand and since they do not consider the diffusion of the molecule in space, they are not suitable to investigate the hybridization of target and probe under microarray conditions.

## III. THE BEAD-PIN MODEL

The model we develop is closest in spirit to the "bead" models (Fig. 1). We model single-stranded nucleic acid chains as linear polymer chains in which each monomer comprises a bead rigidly attached to a pin. The bead represents the sugar molecule, while the pin represents the nitrogenous base. We model phospho-diester bonds as rigid rods that connect two consecutive sugar beads and form, with the beads, the backbone of the chain.

The sugar beads sit on the nodes of a three-dimensional triangular lattice (Fig. 2). This lattice is commonly used in simulations of polymers—see, for instance, Ref. [57]—because: *(i)* each node has a larger number of first neighbors than cubic lattices, implying that a greater number of symmetries are preserved [58], and *(ii)* it is not possible for two strands to cross—a situation that is almost unavoidable for cubic lattices in which movements of the beads to next-nearest neighbor nodes are allowed [59]. Note that at each time, we allow a single bead to sit on any lattice site.

### A. Lattice Configurations

To model the stiffness of the chain, we restrict the angle between two consecutive bonds (Fig. 2b). The model generates sequence-dependent elastic properties [60] by means of base spe-cific stacking interactions [15, 61] (Fig. 3). Because bonds that link two consecutive sugars in the strand can rotate almost freely [60], we impose no restrictions on the direction of the base pins.

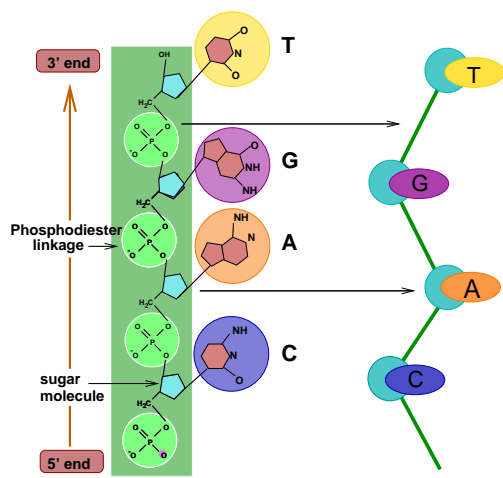

FIG. 1: (Color online) Meso-scale representation of the basic "units" comprising a nucleic acid chain: Phospho-diester bonds (green circles), sugar molecules (light blue pentagons) and nitrogenous bases (large colored circles). The diagram to the right illustrates the different units in our model: Sugar molecules (blue circles) are bonded by phosphates (green straight lines) to form the phosphate backbone of the nucleic acid (green box); colored pins represent the nitrogenous bases. Here and in the following figures, we use the following color coding: yellow stands for Thymine (T), purple stands for Guanine (G), orange stands for Adenine (A) and dark blue stands for Cytosine (C).

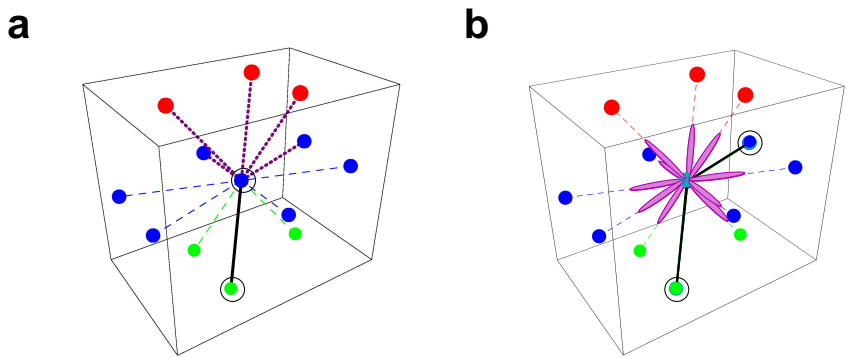

FIG. 2: (Color online) Lattice imposed constraints for the nucleic acid chain configurations. **a,** We use a three-dimensional triangular lattice. We constrain consecutive phospho-diester bonds to have an angle larger or equal to $60^o$ in order to mimic the stiffness of the sugar-phosphate backbone. The diagram illustrates the conformations allowed for two consecutive bonds in a three-dimensional triangular lattice. The black solid line and the black circles represent the reference bond and beads, respectively, and the purple dashed lines indicate the allowed conformations for the following bond. In the diagram, colored dots represent lattice sites which are nearest neighbors of the central blue dot. Different colors indicate the plane on which the site sits: top (red), middle (blue) and bottom (green). **b,** The phospho-diester bond can be easily torqued, hence a sugar-base complex can take any spatial orientation provided it does not overlap with the phospho-diester bond. The diagram illustrates the ten possible orientations that a base (pin) can take (purple ellipses) for a given conformation of the polymer chain indicated by the black circles and black solid lines.

An important factor concerning the implementation of the model is that the characteristic timescale for the rotation of the nucleosides about the chain axis is of the order of nanoseconds [62–65], at least two orders of magnitude faster than the timescale associated with the motion of the monomers in the polymer chain itself, which for molecules tens of bases long is of the order of fractions of microseconds for ssDNA [66] or microseconds for dsDNA [67]. The implication for the dynamic rules implemented in the model is that after the translational motion of a nucleotide, the nucleotide conformation is immediately relaxed to the temperature-specific equilibrium conformation [89]. It follows that the time resolution of our model is finite, hence phenomena taking place at timescales shorter than nanoseconds cannot be investigated.

### B. Interactions

The model allows for different types of pairwise interactions including nucleotide-nucleotide and nucleotide-solvent. These interactions are assumed to be short-ranged and thus restricted to elements occupying neighboring sites in the lattice. In the following, we only describe interactions between pairs of bases. Solvavility effects due to salt concentration can be effectively introduced by changing the values of the interactions.

We consider two types of *nucleotide-nucleotide* interactions: complementarity interactions and stacking interactions. Complementarity interactions lead to Watson-Crick (WC) pair formation through hydrogen bonds. These interactions occur when the pins of a pair of neighboring nucleotides point to one another. Thus, complementarity interactions are not possible between consecutive bases in a strand; although they are possible between bases belonging to the same strand as long as the rigidity conditions described in Fig. 2 are not violated. We show in Fig. 3a the strength of these interactions.

The stacking interaction arises from the fact two bases "like" to "lie" on top of each other. In our model, two consecutive nucleotides are stacked when the pins are parallel to each other and the relative angle between the pin orientation and the phospho-diester bond connecting the two nucleosides is greater or equal than $60^o$. In general, this interaction is stronger for purines than for pyrimidines because of their larger size [60, 61]. However, the strength of the interaction also depends on the sequence and, in the case of dsDNA, on the existence or not of base pairs above and below the considered one. In such a case, opposite bases belonging to adjacent bonded base pairs can be cross stacked (Fig. 3).

The strength of the different interactions shown in Fig. 3 was obtained mostly from experimental data. As a first approximation we consider the interaction to be symmetric, i.e., there is no difference between 5' to 3' and 3' to 5' interactions.

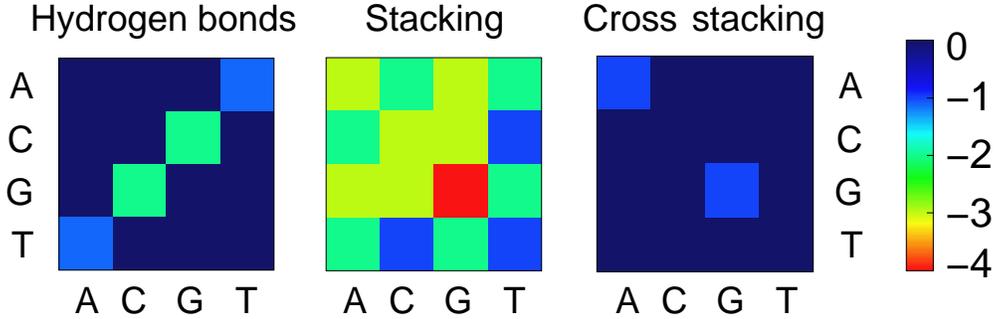

FIG. 3: (Color online) Interaction energies (in arbitrary units) between pairs of nucleotides. We use the color code shown on the right to represent interaction strengths. We obtained the single-strand stacking energies from experimental data reported in Ref. [60] (Chap.8), and the hydrogen bonding and cross-stacking energies from the duplex stacking enthalpies used in the Turner Rules [60] (Chap. 8).

We obtain the base-stacking interactions directly from the enthalpies measured from the thermodynamic parameters for single-strand stacking reported in Chap. 8 of Ref. [60]. For convenience, we rescale them into the range [-4,0] for convenience. Note that the data is incomplete since there are no experimental measurements for the stacking enthalpies for Poly(G) or the combinations AC, GC, AG , GT and CT. To assign the remaining stacking interactions we use the following assumptions: *(i)* Purines have stronger stacking interactions; *(ii)* Gs have stronger stacking interactions than As and Cs have stronger stacking interactions than Ts or Us. The rational for *(i)* is purine's larger size, while the rational for *(ii)* is the greater stability of duplexes comprising G-C bonds.

We compute the hydrogen-bonding and cross-stacking interactions from the duplex stacking enthalpies used in the Turner rules. As a first approximation, we consider that cross-stacking interactions only occur between purines. As for the base-stacking interactions we rescale all the interaction values into the range [-4,0].

### C. Chain motion

A major challenge when modeling the kinetics of lattice polymer chains is the implementation of thermal dynamics that: *(i)* sample the whole phase space, *(ii)* reproduce thermodynamic equilibrium properties, and *(iii)* are realistic and consistent with the kinetic features of the system being modeled. The selection of realistic chain movements that preserve ergodicity and do not introduce spurious symmetries into the conformations of the polymer is, thus, of the greatest importance.

In the past, there has been some discussion on whether the use of Monte Carlo (MC) dynamics is a valid tool to investigate polymer kinetics, since it was initially formulated to investigate static equilibrium properties. There is, however, plenty of examples in polymer litterature showing that by choosing an appropriate set of rules of motion and the correct simulation time scale, the results obtained using MC dynamics are as reliable as those obtained with molecular dynamics [68]. In fact, there is evidence that simulations using MC sampling reproduce the dynamics observed experimentally [68, 69].

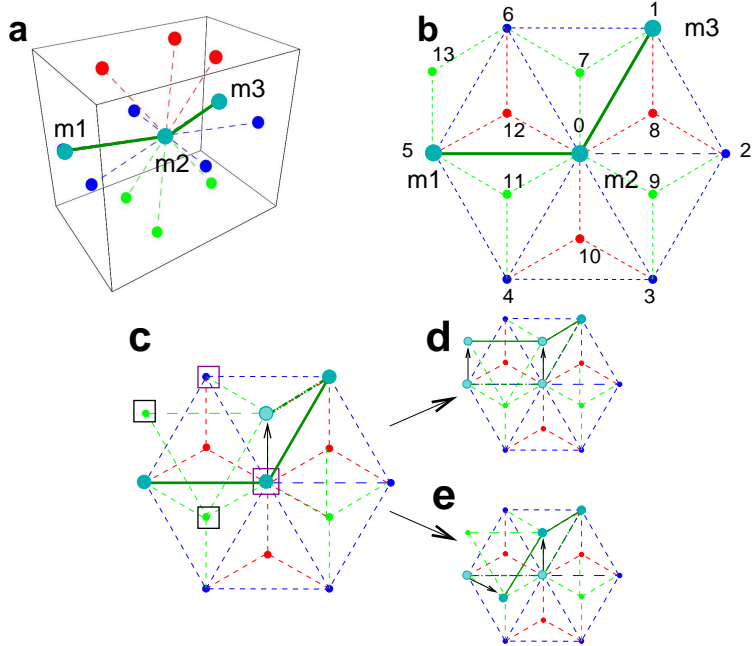

FIG. 4: (Color online) Chain motion on the lattice. In the panels, different colors indicate the different planes to which the sites belong: Blue for the central plane, red for the plane above and green for the plane below. We label the central site "0" and we number the twelve neighboring sites from one to twelve. Site "13" is an example of a next nearest neighbor of the central site "0" that is a nearest neighbor of sites "5" and "6". **a,** Initial configuration of a polymer chain comprising three monomers ($m_1, m_2, m_3$) sitting on the nodes of a three-dimensional lattice. To illustrate our algorithm for the motion of the chain, we consider the motion of the monomer $m_2$. **b,** Projection onto the central plane of the polymer configuration and the neighboring lattice sites. The color code is the same as in (a). Monomer $m_2$ can move with equal probability to any of the ten empty neighboring sites. **c,** $m_2$ moves to site "7" (indicated by the black arrow). This site is a nearest neighbor of site "1" in which $m_3$ sits, but it is not a nearest neighbor of site "5" which is occupied by $m_1$. Since consecutive monomers in the polymer chain must occupy neighboring sites in the lattice, $m_1$ must "reptate", i.e., it must move to a neighboring site which is also a neighbor of site "7". These are sites {"0", "6", "11", "13"} indicated by boxes. Purple boxes show sites which cannot be occupied {"0", "6"}, because the final configuration would violate stiffness constraints. Black boxes indicate acceptable sites {"11", "13"}. With equal probability, monomer $m_1$ can move to either of the two acceptable sites. The two possible final configurations are shown in panels **e** and **d**.

A number of algorithms using MC dynamics have been proposed overtime to investigate polymers. One of the most popular is the Verdier-Stockmayer model [70], in which a number of local moves can be performed depending on the local conformation of the monomers: namely, the so-called "crankshaft", "end-bond", and "kink-jump" movements. MC simulations using these dynamics have been shown to reproduce some real kinetic properties of homopolymers and pro-teins [59, 69, 70]. However, this algorithm has problems in the sampling of phase space [59]. Specifically, the relaxation of kinks toward the center of the polymer chain is very slow and the polymer can get locked in some configurations.

With other chain "moves", such as reptation (or "slithering snake") [71] and "pivot relaxation" [72], the sampling of phase space is much improved and the relaxation toward equilibrium is much faster. However, the rules of motion proposed in these algorithms are not "realistic" moves that happen in real polymers under dilute conditions [59]. Nevertheless, by constraining the reptation to a number of selected internal monomers, a modified reptation algorithm can be used to "propagate kinks" along the chain while keeping a correct description of the kinetic properties [73, 74].

In our model, we use a generalized version of this "internal reptation" model which includes, but is not restricted to, the propagation of kinks along the polymer (Table I). Our dynamics includes all the local movements considered in the Verdier-Stockmayer algorithm, as well as the propagation of "local deformations" along the chain. This generalized dynamics has the advantage that in order to generate a new configuration, one does not have to study the local configuration of the monomers to see which local movement is possible as it happens in the Verdier-Stockmayer algorithm. The only constrain for the new configuration is that the stiffness conditions be fulfilled.

Because the timescales for the motion of the entire polymer and for the rotation of the pins differ by a factor of a thousand, we use a modified MC scheme that considers separately the translational motion of the beads and the rotational motion of the pins. Specifically, we "split" the motion of the nucleotide chain into two steps: *(i)* bead motion and *(ii)* pin motion. Table I describes in detail the algorithm by which we implement the motion of beads and pins—note that the pins thermalize regardless of whether the beads change their configuration or not.

In the algorithm described in table I, we consider single strands whose bases can form WC pairs. If we consider two single strands bonded to each other, an immediate extension of the algorithm is to consider the simultaneous motion of the two bonded chains.

## IV. MODEL VALIDATION

In order to validate our model, we first perform basic tests that ensure that our model displays a physically sound behavior. Next, we study static and kinetic properties of ssDNA hairpins, which are self-complementary single strands linked by a non-complementary loop (Fig. 5). Hairpin conformations are ubiquitous in nature. In RNA, they dominate the secondary structure and are responsible for the fast folding into the native structure [11], while in DNA they are involved in important biological processes such as the regulation of gene expression [14, 53] and DNA recombination [12, 75] and transposition [13, 76]. Importantly, hairpins are not static structures: In thermal equilibrium, they fluctuate between open and closed states (Fig. 5), providing with an ideal model system for the investigation of single-strand properties.

| 1 | Randomly select a monomer, $m_i$. | | | |
|---|---|---|---|---|
| 2 | If $m_i$ is not bonded to its own chain:. | | | |
| | 2.1 | List all the empty neighboring sites that fulfill the stiffness constrains with at least one of the sites occupied by the two neighboring monomers in the chain $m_{i-1}$ and $m_{i+1}$. | | |
| | 2.2 | Randomly select a new site from the list for the selected monomer to move to. | | |
| | 2.3 | Select the direction of the chain that will reptate: | | |
| | | 2.3.1 | If the new position of the selected monomer is a neighbor of only the site occupied by the left (right) neighboring monomer in the chain, the monomers toward the right (left) end of the chain reptate. | |
| | | 2.3.2 | If the selected site is a neighbor of the sites occupied by both $m_{i-1}$ and $m_{i+1}$, then: | |
| | | | 2.3.2.1 | If the new configuration satisfies stiffness constraints: |
| | | | | 2.3.2.1.1 With probability $p$, only the selected monomer moves to the selected site. No other monomers reptate. |
| | | | | 2.3.2.1.2 With probability $1-p$, select randomly one of the ends (right or left), and allow the monomers along the chain toward the selected end to reptate. |
| | | | 2.3.2.2 | If the new configuration does not satisfy stiffness constraints with the left (right) neighboring monomer in the chain, allow the monomers along the chain toward the left (right) end of the chain reptate. |
| | 2.4 | Move the selected monomer to the selected site and iterate the following steps for the monomers between the selected monomer and the selected end. | | |
| | | 2.4.1 | For the following monomer, build a list of empty neighboring sites that satisfy stiffness conditions with the previous monomer in the chain. | |
| | | 2.4.2 | Select randomly a new site from the list for this monomer to move to. | |
| | | 2.4.3 | If the selected site is a neighbor of the sites occupied by the two neighboring monomers in the chain and the new configuration satisfies stiffness conditions, stop the reptation with probability $p$. | |
| 3 | If $m_i$ is bonded to its own chain, propose a change of orientation. Accept or reject the change using the Metropolis algorithm. | | | |
| | 3.1 | If the change is accepted, follow the procedure described in step 2. | | |
| | 3.2 | If the change is rejected, the pair of bonded monomers move simultaneously in the same direction while the remaining monomers remain in the previous position. | | |
| | | 3.2.1 | List the pair of neighboring sites of the pair of bonded monomers that are neighbors of the consecutive monomers in the chain and satisfy stiffness conditions. | |
| | | 3.2.2 | Select randomly a pair of sites among the list and perform the movement. | |
| 4 | Verify that none of the pin orientations overlaps with the backbone. | | | |
| | 4.1 | If a pin orientation overlaps with the backbone, select randomly a new orientation for that pin. | | |
| 5 | Compute the energy of the new configuration. | | | |
| 6 | Accept or reject the change in configuration using the Metropolis algorithm. | | | |
| 7 | For a number of times equal to the number of monomers in the chain, repeat the following steps: | | | |
| | 7.1 | Randomly select a monomer $m_j$. | | |
| | 7.2 | Randomly select a new orientation for the pin of $m_j$ that does not overlap with the backbone of the polymer. | | |
| | 7.3 | Compute the new energy. | | |
| | 7.4 | Accept or reject the change in orientation using the Metropolis algorithm. | | |

TABLE I: Algorithm for the motion of a single chain.

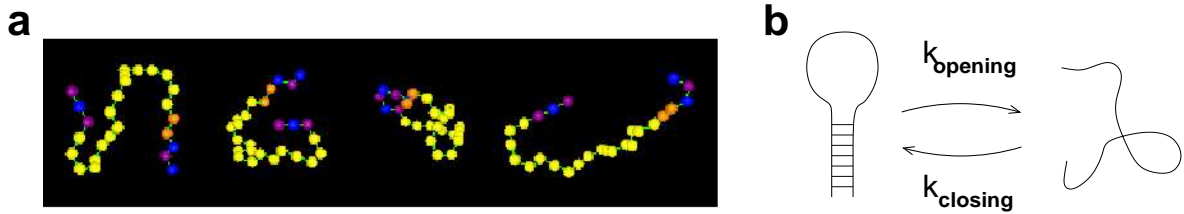

FIG. 5: (Color online) Single-strand DNA hairpins. **a,** Sample configurations of a ssDNA hairpin, comprising 25 nucleotides with sequence GCGTT-$T_{15}$-AACGC, on a three-dimensional triangular lattice. Spheres with different color indicate nucleotides with different bases: Orange for adenine (A), purple for guanine (G), yellow for thymine (T), and blue for cytosine (C). Note that the lattice symmetries are almost unnoticeable. **b,** Schematic illustration of the transition between open and closed states for a hairpin loop. The hairpin switches between open/coil and closed/native states with characteristic rates $k_{\text{opening}}$ and $k_{\text{closing}}$.

### A.  Sampling of configuration space

First, we test if the motion algorithm implemented in our model is ergodic [77]. To this end, we investigate the sampling of configuration space for a polymer chain moving according to the algorithm described earlier and for different values of $p$. We study two polymers comprising 6 and 8 monomers at infinite temperature. Our results indicate that the sampling of configuration space becomes more uniform as $p \to 0$ (Fig. 6). Importantly, our analysis also suggests that for $p$ as large as 0.1, the sampling of configuration space is already essentially uniform. This is of practical relevance because even for $p$ of order 0.1, one already observes a substantial decrease in simulation times. This decrease arises from the fact that in MC simulations, the energy difference between configurations increases with increasing number of moving monomers. Larger energy differences make it less likely for the move to be accepted, resulting in longer equilibration times.

### B.  Average radius of gyration

The radius of gyration $R_g$ is the mean distance of all monomers to the center of mass of the polymer,

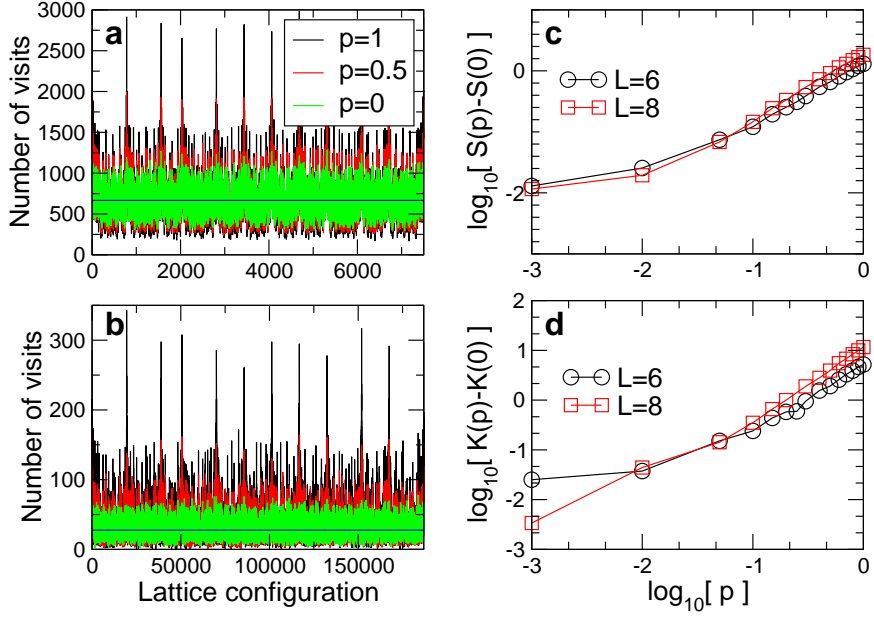

FIG. 6: (Color online) Sampling of the configuration space. Sampling of the configuration space at infinite temperature for polymers comprising **a,** six and **b,** eight monomers. Different color lines represent different values of the probability $p$. At high temperature, one expects all configurations to be sampled with equal rates (whose value is shown by the blue line). The average rate is the number of time steps in the simulation divided by the total number of configurations for the polymer. The number of configurations for a polymer with six (eight) monomers that sits on a three-dimensional lattice and satisfies the stiffness constraints indicated in Fig. 2 is 7,500 (186,792). For each polymer size, we collected statistics for 5,000,0000 time steps. Our results demonstrate that the polymer samples conformation space more uniformly for smaller values of $p$. **c,** Skewness $S(p)$ of the distribution of sampling rates of the conformation space of the polymer for different values of $p$. The skewness measures the asymmetry of the distribution. For perfect sampling, we expect the distribution to be normal, that is $S = 0$. For $p = 0$, we find $S = 0.35$ ($L = 6$) and $S = 0.59$ ($L = 8$), in good agreement with this expectation. **d,** Kurtosis $K(p)$ of the distribution of sampling rates of the conformation space of the polymer. The kurtosis measures the decay rate of the tails of the distribution. For a normal distribution, one has $K = 3$. For $p = 0$, we find $K = 2.6$ for $L = 6$ and $K = 3.3$ for $L = 8$, in good agreement with this expectation. Note that both $S$ and $K$ take smaller values for $p < 0.01$ for the longer polymer. This suggests that as the length of the polymer increases, the differences in the distributions for small $p$ with respect to $p = 0$ become smaller.

$$R_g = \frac{1}{L}\sum_{i=1}^{L}\sqrt{\left(\vec{R} - \vec{r}_i\right)\cdot\left(\vec{R} - \vec{r}_i\right)}, \quad \vec{R} = \frac{1}{L}\sum_{i=1}^{L}\vec{r}_i, \tag{1}$$

where $L$ is the number of monomers in the chain.

According to polymer theory [79], the average radius of gyration $\langle R_g \rangle$ scales with the polymer length (i.e., the number of monomers) as $\langle R_g \rangle \sim L^\nu$, with $\nu > 0$. In our model, we have included volume constraints since a lattice site cannot be occupied by more than one monomer simultaneously, but we also have stiffness constraints and therefore we should expect to obtain an exponent value somewhat larger than the value for a self-avoiding random walk $\nu = 0.6$. Figure 7a shows that our simulations agree with theoretical expectations, since the average radius of gyration for different polymer lengths scales as $\langle R_g \rangle \sim L^\nu$, with $\nu = 0.75 \pm 0.07$.

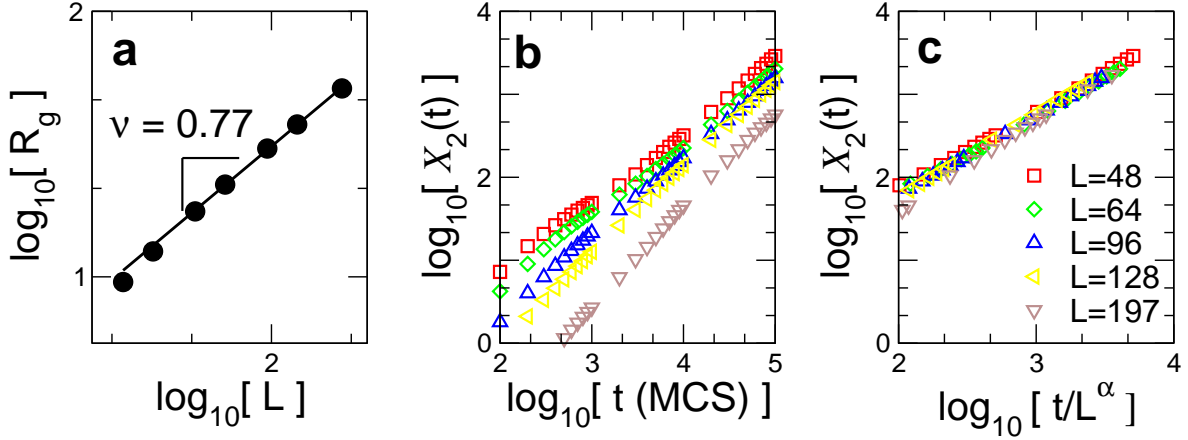

FIG. 7: (Color online) **a,** Average radius of gyration $\langle R_g \rangle \sim L^\nu$ versus polymer length $L$. The solid line indicates the best fit to the expected power law behavior $\langle R_g \rangle \sim L^\nu$, obtaining $\nu = 0.75 \pm .07$. **b,** Mean squared displacement $X_2(t)$ versus time for different polymer lengths $L = 48$–$197$. Note that $X_2$ grows linearly with time as expected in a diffusive process [78]. **c,** By scaling the data in (b) by $D \propto L^\alpha$ with $\alpha = 0.80 \pm 0.13$, we are able to collapse all the data onto a single curve. The data displayed in the plots are averages over 5000 runs 100,000-150,000 time steps long using $p = 0.05$.

### C. Diffusion

A polymer comprising $L$ monomers diffuses with a diffusion constant that scales with the length of the chain $D \sim L^{-\alpha}$ [78]. In order to test this prediction, we measure the mean squared displacement $X_2(t)$ of the center of mass of the chain as a function of time,

$$X_2(t) \equiv \langle [\vec{R}(t) - \vec{R}(0)]^2 \rangle , \qquad (2)$$

where $\langle ... \rangle$ indicates the averages over different dynamical histories of the chain, $\{\vec{r}_i\}$ is the set of positions of the monomers, and each time step corresponds to a single chain movement. In the diffusive regime, $X_2$ scales linearly with time: $X_2 \sim D\,t$.

We study the mean squared displacement versus time for polymers of lengths $L = 48$ to $L = 197$. We find that the linear regime is reached after approximately a hundred time steps. This linear growth is apparent in Fig. 7b. By scaling all the curves for the different polymer lengths, we find that the diffusion coefficient scales as $D \sim L^\alpha$, with $\alpha = 0.80 \pm 0.13$.

### D. Nucleotide movements: Thermal dynamics

To test whether at finite temperatures our model samples the different configurations with Boltzmann statistics, we study the simplest hairpin structure possible: A-TTTT-T. This hairpin, which comprises a one base-pair stem and a four-base T-loop, is the simplest because T has the weakest

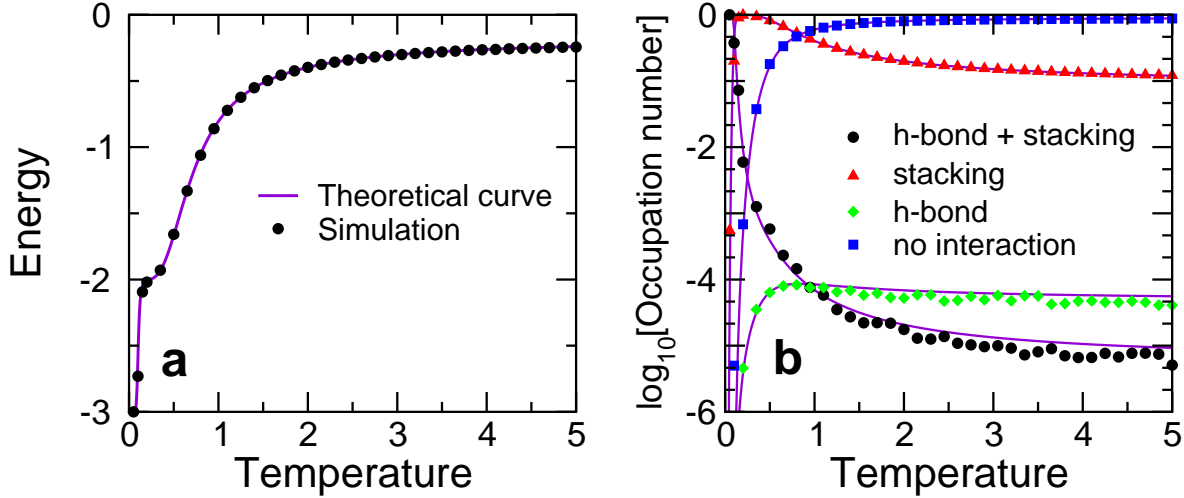

FIG. 8: (Color online)**a,** Equilibrium energy of a hairpin with sequence A-TTTT-T. This hairpin has an internal loop comprising 4 T's and a stem with a single base pair A-T. Differently from Fig. 3, we consider that the stacking energy of the Ts is zero. Under these conditions, a hairpin configuration can only take four energy values, $\epsilon = -3$, when the hairpin is closed,(A-T hydrogen bond is formed) and A is stacked with its neighboring T; $\epsilon = -2$, when the A is stacked with its neighboring T and the hairpin is open; $\epsilon = -1$, when the hairpin is closed but there is no stacking between the A and its neighboring T; $\epsilon = 0$, for all other cases. Under these conditions, the exact number of configurations for each energy level, $g_i$ can be computed. **b,** Occupation number $\langle n_i(T) \rangle$ of each energy level, $i = 0 - -3$ as a function of temperature. Colored dots indicate the numerical results obtained from averages over 5,000,000 Monte Carlo steps (MCS) using the parallel tempering MC method [80]. Purple solid lines correspond to the theoretical expressions for the energy $e = \sum_i \epsilon_i \, g_i \, e^{-\epsilon_i/T}/\mathcal{Z}$ in (a), and the occupation number $\langle n_i(T) \rangle = g_i \, e^{-\epsilon_i/T}/\mathcal{Z}$ in (b), where $\mathcal{Z} = \sum_i g_i \, e^{-\epsilon_i/T}$ is the partition function. Note the excellent agreement between theoretical predictions and simulation results.

interactions of all nucleic bases.

The question of the uniform sampling of all possible configurations for very large temperatures was already addressed in Sec. IV A. We now calculate the equilibrium energy of the hairpin as a function of temperature (Fig. 8a). To simplify the calculations, we set the stacking interactions of the Ts to zero. Under these conditions, there are two possible interactions with non-zero energy: the formation of the A-T WC pair in the stem (energy $\epsilon = -1$) and the stacking interaction between the A and its neighboring T (energy $\epsilon = -2$). Therefore, there are four possible energy values $\epsilon = -3, -2, -1, 0$. The minimum energy $\epsilon = -3$ corresponds to a closed hairpin with A and T stacked. If the hairpin is open but A and T are stacked, the energy is $\epsilon = -2$. If the hairpin is closed but there is no AT stacking, the energy is $\epsilon = -1$. In all other cases, the energy is 0.

By enumerating the possible pin conformations for each of the the 7,500 different lattice configurations for a polymer with $L = 6$, we are able to compute the degeneracy of each energy level: $0, -1, -2$ and $-3$. Hence, we can calculate the expected occupation numbers—i.e., average population—of each energy level as a function of temperature (Fig. 8b). At high temperatures, one expects the energy to be dominated by the configurations with zero energy, because of their large number, while at lower temperatures, one expects that the dominant contribution comes from those configurations with lower energies. As Figure 8 demonstrates, we find excellent agreement between the simulations and the theoretical predictions [90].

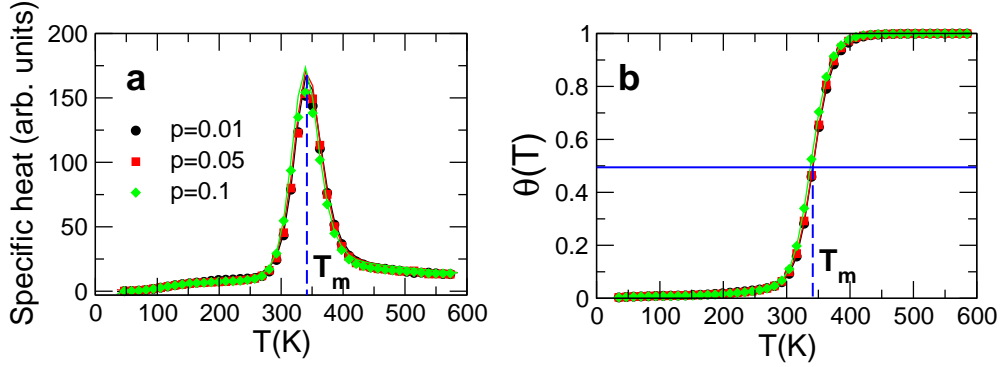

FIG. 9: (COlor online) Test of the equilibrium properties of a hairpin with sequence GGATAA-T$_4$-TTATCC. We performed simulations using the parallel tempering method [80] for the cases $p = 0.01$, $0.05$ and $0.1$. Results correspond to averages over 10,000,000 MCS. Simulation temperatures in the range [0,1] were mapped into absolute temperatures using the conversion factor $T_m^{Sim}/T_m^{MFOLD} = 1.71 \times 10^{-3}$ obtained in Fig. 10 for the ionic conditions [Na$^+$]=1M and [Mg$^{++}$]=0 M. **a,** Specific heat as a function of temperature calculated as (i) the derivative of the energy with respect to temperature $dE/dT$ (solid lines), and (ii) $c = (\langle E^2 \rangle - \langle E \rangle^2)/T^2$ (symbols) [77]. At equilibrium, fluctuation-dissipation relations must be fulfilled and the two methods must lead to equal estimates. This is indeed what we observe. Furthermore, note that the agreement between the two methods is excellent even in the melting region when the heat capacity has its peak. We also checked that at low temperature the hairpin reaches its minimum energy configuration. Thus, this test demonstrates that we reach equilibrium in our simulations and that the equilibrium properties that we measure are correct. **b,** Melting curve for the values of $p$ considered in (a). Note that the curves are insensitive to the specific value of $p$ in the range considered. As expected, the fraction of broken bonds in the stem goes from zero at low temperatures (where the low energy of the closed/native state dominates the partition function), to one at high temperatures (where entropy dominates). Blue dashed lines indicate the melting temperature in both plots: At the specific heat peak, and at the point where $\theta = 0.5$. We obtain for the two cases $T_m = 341 \pm 3$ K.

### E. Melting temperatures

In order to show that our model correctly describes hairpin properties, we test whether equilibrium properties such as melting temperatures and closing times are in agreement with experimental observations. First, we demonstrate that our model is able to reach equilibrium and that we do observe a transition from a high temperature region dominated by open configurations to a low temperature region dominated by closed configurations (Fig. 9). We measure the average fraction of broken bonds $\theta(T)$ as a function of temperature and find a typical melting curve that goes from one at high temperatures, where hairpins are mostly open, to zero at low temperatures, where open configurations dominate the partition function. The temperature at which $\theta(T_m) = 0.5$ defines the

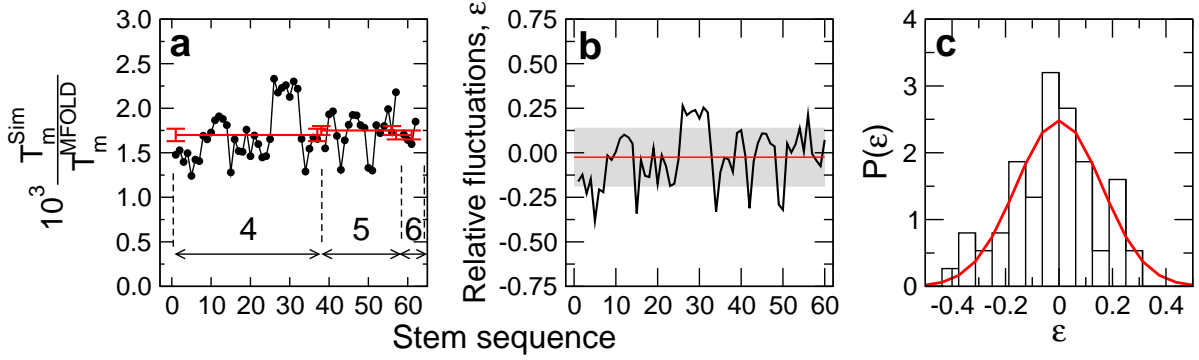

FIG. 10: (Color online) Interaction Testing: Comparison of results from simulations with theoretical values. **a,** Ratio of the melting temperature estimates, $r \equiv T_m^{Sim}/T_m^{MFOLD}$, obtained with our model and with M. Zucker's DNA folding server [1, 81] for different hairpins with stems of 4, 5 and 6 bases long. For all hairpins, the loop comprises four Ts. Melting temperatures from the folding server correspond to the following ionic concentrations: [Na$^+$]= 1 M and [Mg$^{++}$]=0 M. Note that in the interaction tables of Fig. 3, we do not consider any difference between oligonucleotides starting at the 5' or the 3' end. The reason for this modeling choice is that the fluctuations observed in the melting temperature obtained in each case: 5' to 3' ($T_m(5')$) and 3' to 5' ($T_m(3')$) sequences were considerably smaller than the fluctuations of the whole data set. Therefore, the melting temperature used in the data shown corresponds to the average of both $T_m$s. The red solid lines correspond to the average factors for hairpins with stems comprising 4, 5 and 6 base pairs. Note that the fluctuations are quite small. **b,** Relative fluctuations $\epsilon = (r - \bar{r})/r$ of the ratio of temperatures with respect to the mean ($\bar{r} = 1.71 \times 10^{-3}$). The red line represents the mean $\bar{\epsilon} = -0.025$ and the gray band represents the region within one standard deviation of the mean. **c,** Normalized distribution of $\epsilon$. The red line corresponds to a Gaussian fit of zero mean and standard deviation $\sigma = 0.15$. We obtained the melting temperatures from parallel tempering Monte Carlo simulations for temperatures in the range 0.06 to 1, and performing averages over $3 \times 10^6$ $5 \times 10^6$ and $10 \times 10^6$ MCS for hairpins with stems comprising 3, 4, 5 and 6 base pairs, respectively. All data corresponds to the case $p = 0.05$, but we found no significant changes for different values of $p$. The analysis of different ionic conditions yields similar fluctuations but different conversion factors, suggesting that salt concentration and temperature play a similar role in our model.

melting temperature, which is also the temperature at which the melting curve has an inflection point and the specific heat has a maximum [91]

Next, we investigate if the values for the nucleotide interaction energies that we derived from experimental data in Ref. [82] lead to "self-consistent" predictions of melting temperatures of hairpins with different sequences. We perform simulations for more than 60 hairpins with randomly sampled stem sequences with stems comprising 4, 5 and 6 base pairs and loops comprising four Ts. We show in Fig. 10a the factor necessary to convert the melting temperatures in our simulations $T_m^{Sim}$ into the the melting temperatures obtained using the $T_m$ server of M. Zuker [1, 81]. It is visually apparent that we obtain an approximately constant conversion factor for all those hairpin sequences.

In order to better evaluate the fluctuations of the conversion factor, we show in Fig. 10b the relative fluctuations of the sequence specific conversion factor to the average conversion factor. Note that most cases are within 30% of the average, and that the standard deviation is only 15%. More-

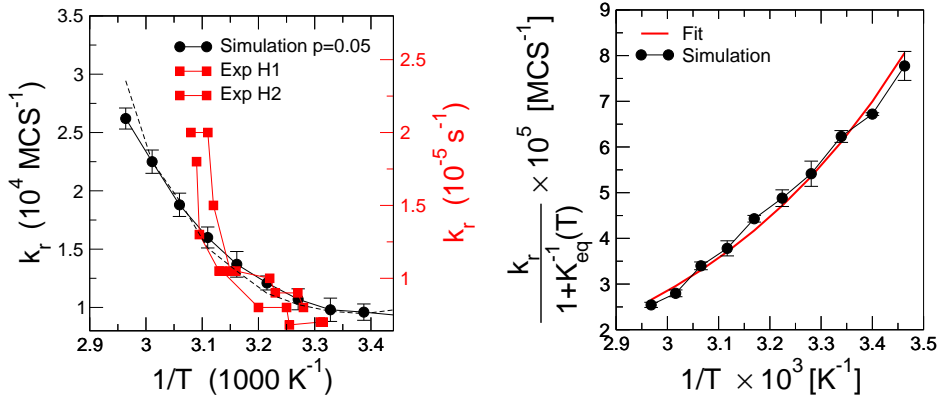

FIG. 11: (Color online) Kinetic properties. **a,** Relaxation rates $k_r$ for the hairpin GGATAA-$T_4$-TTATCC versus inverse temperature. Black circles correspond to the data obtained from the simulations while red squares correspond to experimental data obtained by Ansari et al. with the same hairpin for two different experimental setups [16, 18]. The dashed line corresponds to the fit to a two state model with Arrhenius dependence of the relaxation rates; cf. (b). Our data were obtained by averaging over different dynamical histories (~200) for temperatures in the range $T_m \pm 0.07\, T_m$. We run simulations with a wide range of MCS (150,0000 to 1,800,000) to make sure that our estimates of the relaxation rates converged. Simulation temperatures have been rescaled by a factor $1.82 \times 10^{-3}$ to convert the temperature into Kelvin. Note that we use a different factor from that obtained in Fig. 10 to adjust to the experimantal conditions used in Ref. [16]. In order to convert simulation rates to experimental rates, one needs to use a factor of approximately $10^{-9}$. This value suggests that the timescale of a single MC step is approximately one nanosecond. **b,** Two state analysis. $k_r/(1 + K_{eq}^{-1}(T))$ (black dots) versus inverse temperature. $K_{eq}$ is known from equilibrium measurements (Fig. 9) as $K_{eq}(T) = 1/\theta(T) - 1$. The solid red line corrsponds to the fit to the expression in Eq. (4). The two fitting parameters are $E_a$, the activation energy, and $A$, a phenomenological constant rate. The parameters for the best fit are $E_a = -3.5 \pm 0.4$ kcal/mol and $A = 30 \pm 2$ s$^{-1}$.

over, as shown in Fig. 10c, these fluctuations are well described by a Gaussian distribution with zero mean and standard deviation 0.15, indicating that there is no apparent bias in our estimation of melting temperatures [92]

We checked that the fluctuations of the ratios between the experimental melting temperatures for hairpins with short stems reported in [60] and the values predicted by the server are ten times smaller than the fluctuations observed with the ratios between the simulation results and the server predictions.

### F. Relaxation rates

To validate the dynamics of our model, we compare the kinetic measurements obtained from simulations at a fixed temperature with experimental results. Specifically, we measure the relaxation rates $k_r$ for a hairpin of sequence GGATAA-$T_4$-TTATC which was studied experimentally in [16]. The relaxation rate is defined as

$$k_r = 1/\tau_{\text{closing}} + 1/\tau_{\text{opening}}, \tag{3}$$

where $\tau_{\text{closing}}$ and $\tau_{\text{opening}}$ stand for the closing and opening times, respectively. In figure 11a, we show that simulation results for $k_r$ (black dots) are in agreement with experimental measurements (red dots) and show a decrease in rate with $1/T$. Note that in order to convert simulation rates to experimental rates, one needs to use a factor of the order of $10^{-9}$ s/MCS. This value suggests a correspondence between one Monte Carlo step and one nanosecond, which is the timescale at which nucleosides move and get thermalized [65]. Recall that in our algorithm nucleosides are thermalized within one MCS, which is thus consistent with the experimental timescales.

The simplest description for the folding/unfolding transitions of a hairpin is a two-state system (open and closed) with a transition state at a constant energy barrier $E_a$. Two-state models are commonly used to describe the kinetics of the unfolding of single-domain proteins and hairpins [16, 20, 83–85]. Within this description, that we suppose to be valid close to $T_m$, the relaxation constants are assumed to have an Arrhenius dependence on the barrier. In this scenario, the relaxation constant is described by

$$k_r = A \frac{e^{-E_a/T}}{1 + K_{eq}}, \qquad (4)$$

where $K_{eq}(T) = 1/\theta(T) - 1$ is the equilibrium constant, $T$ is the absolute temperature (we have set $k_B = 1$ for convenience), and $A$ is a phenomenological constant rate. Figure 11b shows that $k_r/(1 + K_{eq}^{-1}(T))$ is well fit by an exponential with a negative activation energy $E_a \approx -3.5$ kcal/mol consistent with the analysis of the experimental data for the same hairpin by Ansari et al. [16]. Negative activation energies are believed to be a hallmark of zipping processes in which the transition state has a lower energy than the coil configuration. In such processes, the rate-limiting step is the formation of a nucleus with a small number of hydrogen bonds—between residues in polypeptides or bases in nucleic acids—that immediately leads to the complete folding of the molecule [60]. This is not unlike the situation found in oligonucleotide dimers [86], protein $\beta$-sheet hairpins [83, 87] and protein $\alpha$-helices [88] [93].

## V. CONCLUSION

Understanding the configurational dynamics of nucleic acids is relevant to many open questions such as the folding of RNA or the hybridization of two separate DNA strands in microarrays. A particularly challenging problem is to understand how base heterogeneity affects mechanical and kinetic properties of nucleic acids at meso-scales.

In order to provide a new window into these questions, in this paper we have described a novel mesoscopic model for nucleic acid chains. The main feature of the model is that it considers single-strand properties individually, making it suitable for the study of double-stranded as well as single stranded-nucleic acids. We have demonstrated, that the dynamical rules implemented are physically sound, and that they are realistic. Specifically, we performed a number of comparisons of static and dynamic properties obtained in our simulations with those for ssDNA hairpins and found good agreement. All these results validate our model making it a suitable tool for the investigation of processes in which single-strand properties are relevant, such as the formation of complex structures such as H-pseudoknots which cannot be predicted by current models.

| 1 | Select a new lattice configuration for the polymer keeping the pin orientations fixed. That is, go through steps 1 to 5 in the algorithm described in (Fig. 6) |
| 2 | Thermalize the nucleotides in this new lattice configuration. That is, change the orientation of the pins following step 7 in the algorithm described in Fig. 6. |
| 3 | Compute the energy of this new global configuration. |
| 4 | Accept or reject this new global configuration according to the Metropolis algorithm. |

TABLE II: Alternative Monte Carlo scheme for the dynamics of a single -stranded nucleic acid.

We thank A. Aalberts, A. Ansari, C. A. Ng, D. B. Stouffer, A. Modragon, and M. Zuker for numerous suggestions and discussions. R.G. and M.S. thank the Fulbright Program and the Spanish Ministry of Education, Culture & Sports. J.W. gratefully acknowledges the support of grants R01-GM054692 and R01-GM058617 from the NIH. L.A.N.A. gratefully acknowledges the support of a Searle Leadership Fund Award and of a NIH/NIGMS K-25 award.

### APPENDIX A: COMPARISON OF DIFFERENT MONTE CARLO SCHEMES

To ensure the validity of the dynamic algorithm described in Sec. III C, a further test is to compare the Monte Carlo scheme that we use to alternative schemes that also consider separately bead and pin motion. A possible alternative is one that we denote scheme B (Table II). This alternative scheme should, in principle, sample configurations with the same equilibrium distribution as the scheme described in Fig. 6. Indeed, Fig. 12 shows that for a hairpin of sequence GGATAA-$T_4$-TTATCC, parallel tempering simulations [80] yield identical melting and energy curves for the two schemes. However, Figure 12b shows that the two schemes are not equivalent when performing simulations at a fixed temperature. In particular, scheme B samples configurations with energies significantly lower than the value obtained with parallel tempering simulations. This observation suggests that scheme B tends to sample the minimal energy conformations of each different lattice configuration, and as a result gets easily "trapped" in local minima. It follows that this alternative scheme is not as good a tool to investigate the DNA dynamics [94].

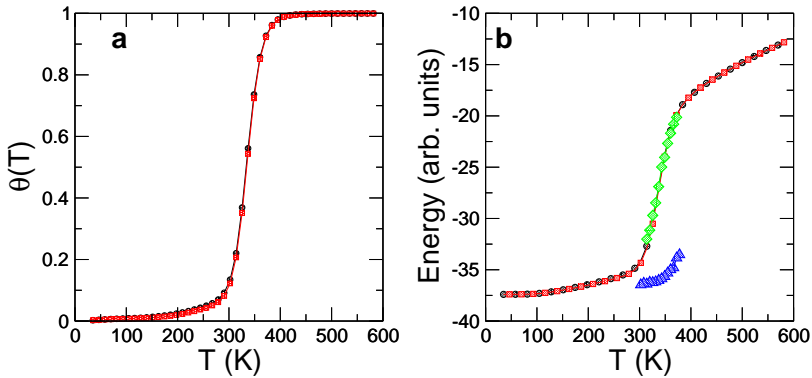

FIG. 12: (Color online) Comparison of two different Monte Carlo schemes for the hairpin of sequence GGATAA-$T_4$-TTATCC. **a,** Melting curve. Fraction of broken bonds $\theta$ with respect to temperature obtained from parallel tempering simulations averaging over $2 \times 10^7$ MCS. Black symbols correspond to the Monte Carlo scheme defined for the model (Table I) and red symbols correspond to results for scheme B (Table II). Note that the curves are practically undistinguishable. **b,** Energy versus temperature. Black and red symbols correspond to parallel tempering results, following the color code in (a). Green and blue symbols correspond to single temperature simulations in the transition region. Each point corresponds to averages over 200 different histories 1,200,000 MCS long. Green symbols correspond to the scheme defined for our model and blue symbols correspond to scheme B. Note that while results for the model scheme show an excellent agreement with the parallel tempering results, results for scheme B show average energies far below the parallel tempering curve.

**APPENDIX B: OPTIMIZATION OF THE INTERACTION PARAMETERS**

To provide better agreement between the simulation melting temperatures and the melting temperatures predicted by M. Zucker's server [1], we need to optimize the interaction strengths presented in Fig. 3 and used in our simulations. To this end, we must define a cost function and select the set of parameters that minimizes it. In our case, an obvious choice for the cost function is the standard deviation of the relative fluctuations of the ratios between simulation and theoretical melting temperatures (Fig. 10). This is, however, a time-consuming task that we have not concluded yet.

To improve the parameter choices, we analyze the melting temperatures for hairpins with stems comprising four base pairs. We find out that hairpins whose stems are rich in GC pairs are more stable than they should be, whereas hairpins with stems rich in AT pairs are less stable than they should be. This suggests that one may reduce the fluctuations of the temperarure ratios by introducing, for instance, the following modifications: *(i)* reducing the GC bond strength, *(ii)* increasing the AT bond strength, and *(iii)* reducing the GC stacking strength. Figure 13 shows the variation of the temperature ratios $r = T_m^{Sim}/T_m^{MFOLD}$ with respect to the nominal case (10)—black dots—for the following cases: *(i)*—blue triangles, *(ii)*—left yellow triangles, *(iii)*—green diamonds, and changing the three parameters simultaneously—red squares. Note that these changes of parameters result in a decrease of the fluctuations, the best choice being the simultaneous change of the strength of AT bonds, GC bonds and GC stacking interactions. To obtain the best set of parameters, one needs to obtain the overall minimum for the variance among all the possible sets of parameters.

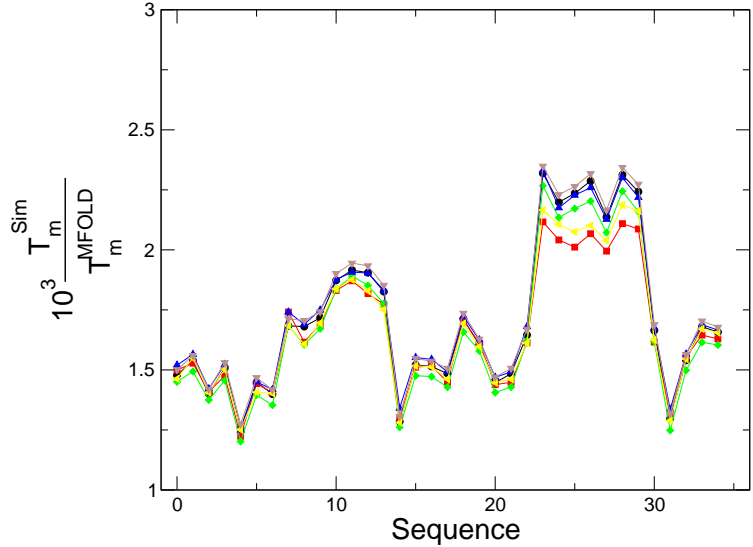

FIG. 13: Interaction optimization. Comparison of results from simulations with theoretical values for different sets of interactions. Ratio of the melting temperature estimates, $T_m^{Sim}/T_m^{MFOLD}$, obtained with our model and with M. Zucker's $T_m$ server [1, 81] for different hairpins with a stem comprising 4 base pairs and a loop comprising four Ts. Different symbols correspond to changing some of the interaction strength values shown in Fig. 3 to which black symbols (○) correspond. Blue symbols (△) correspond to reducing the G-C bond strength. Yellow symbols (◁) correspond to increasing the A-T bond strength. Green symbols (◊) correspond to decreasing the GC stacking strength. Red symbols (□) correspond to performing the three previous changes simultaneously. Note these changes of the parameters result in a decrease of the fluctuations, the minimum variance corresponding to changing the strength of AT bonds, GC bonds and GC stacking interactions at once. We obtained the $T_m$s from parallel tempering [80] simulations averaging over $5 \times 10^6$ MCS, with $p = 0.05$.

[89] This is in apparent disagreement with the motion of real nucleosides which are believed to "adjust" their orientation while the nucleic acid chain moves. Nevertheless, it is still reasonable to assume that at a given instant of time (at which measurements are performed), the different pin conformations are sampled with the equilibrium weights that correspond to the specific lattice configuration of the chain

[90] We also checked that our algorithm is more suitable for kinetic measurements than alternative MC schemes with the same equilibrium distribution (Appendix A).

[91] We also checked that the averages of the different observables at a fixed temperature had converged.

[92] A significant point is that the interaction energies presented in Fig. 3 and used in our simulations, have not yet been optimized to provide the best agreement with Zucker's code [1]. To do this would be a trivial but computationally demanding task (see Appendix B).

[93] However, some experiments in other hairpins have reported positive activation energies [15] a subject which is still under discussion and which might be associated to the experimental technique used, as we shall discuss in another paper.

[94] This observation conveys the point that the parallel tempering technique is useful to recover equilibrium properties, even when the dynamics is not effective at low temepratures, as long as the equilibrium distribution is correct.